\documentclass[12pt,preprint]{aastex}

\usepackage{epstopdf}

\begin{document}

\title{\sc Constraining the Physical Parameters of the
Circumstellar Disk of $\chi$ Ophiuchi}

\author{
C. Tycner,\altaffilmark{1} 
C.~E.~Jones,\altaffilmark{2}
T.~A.~A.~Sigut,\altaffilmark{2}
H.~R.~Schmitt,\altaffilmark{3,4}
J.~A.~Benson,\altaffilmark{5}
D.~J.~Hutter,\altaffilmark{5}
R.~T.~Zavala\altaffilmark{5}}

\altaffiltext{1}{Department of Physics, Central Michigan University,
Mt. Pleasant, MI 48859} 
\altaffiltext{2}{Department of Physics and Astronomy, The University
of Western Ontario, London, Ontario, N6A~3K7, Canada}
\altaffiltext{3}{Naval Research Laboratory, Remote Sensing Division,
Code 7215, 4555 Overlook Ave. SW, Washington, DC 20375 }
\altaffiltext{4}{Interferometrics, Inc., 13454 Sunrise Valley Drive,
Suite 240, Herndon, VA 20171}
\altaffiltext{5}{US Naval Observatory, Flagstaff Station, 10391
W.~Naval Observatory Rd., Flagstaff, AZ 86001}

\slugcomment{Accepted for publication in ApJ}

\begin{abstract}
We present a numerical model describing a circularly symmetric gaseous
disk around the Be star $\chi$~Ophiuchi.  The model is constrained by
long-baseline interferometric observations that are sensitive to the
H$\alpha$ Balmer line emission from the disk.  For the first time our
interferometric observations spatially resolve the inner region of the
circumstellar disk around $\chi$~Ophiuchi and we use these results to
place a constraint on the physical extent of the H$\alpha$-emitting
region.  We demonstrate how this in turn results in very specific
constraints on the parameters that describe the variation of the gas
density as a function of radial distance from the central star.
\end{abstract}


\section{Introduction}

Models of the circumstellar gas around some B-type stars have been
constructed for a number of years.  Because the recombination in the
circumstellar gas produces a series of emission lines, most commonly
in the hydrogen Balmer series, such systems are usually denoted
B-emission or Be stars for short.  Models describing the geometry of
this circumstellar material have varied from spherical
shells~\citep{Gehrz74}, to slabs with constant
thickness~\citep{Kastner89}, to disks with constant opening
angles~\citep{Waters86}, to disks with an exponential density fall-off
in the direction perpendicular to the plane of the
disk~\citep{Marlborough69}.  Although the first models assumed a
constant temperature throughout the disk, more recent models, such as
the ones developed by \citet{Millar98, Millar99}, \citet{Carciofi06},
and \citet{Sigut07}, obtain a self-consistent temperature throughout
the disk by enforcing radiative equilibrium (namely, by balancing the
microscopic heating and cooling rates operating within the disk).

While the temperature structure of Be star disks is now the subject of
sophisticated modeling, the density structure largely remains {\it
ad-hoc\/} with simple, parameterized models. Perhaps the most popular
is a power-law density decrease with radius in the equatorial plane
coupled with the aforementioned exponential drop perpendicular to the
disk. This vertical density structure is assumed to result from the
gravitational equilibrium set by the balance between the gradient in
the gas pressure and the vertical component of the star's
gravitational acceleration. Despite the simplicity of such models,
they have been quite successful in matching a wide range of
observational diagnostics that are sensitive to the geometric
distribution of the circumstellar gas. These include the infrared
excess exhibited by Be stars over normal B stars of the same spectral
class \citep{dou94}, the observed continuum linear polarization of Be
stars~\citep{woo97}, and the shapes of the emission lines observed in
Be star spectra \citep[see for example][]{hum00a}. All of these
diagnostics rely only on spatially unresolved spectra or
spectropolarimetry.  However, even more powerful constraints on the
geometry of the emitting regions in Be stars can be obtained by
combining these spatially {\it unresolved\/} diagnostics with new
optical interferometric observations that directly {\it resolve\/} the
structure of the emitting region on the sky~\citep{Quirrenbach97,
Tycner06b, Gies07}.

To demonstrate how specific constraints can be placed on a disk model
using observations that spatially resolve the source, we chose to
concentrate on the Be star $\chi\;$Oph (B2Ve; HR~6118; HD~148184).
This relatively nearby star with a {\it Hipparcos} distance of $150\pm
17$~pc~\citep{Perryman97} has a strong H$\alpha$ emission and
therefore is an ideal candidate for probing the disk structure at
these wavelengths.

\section{Observations}

\subsection{Interferometry}

\begin{figure}
\plotone{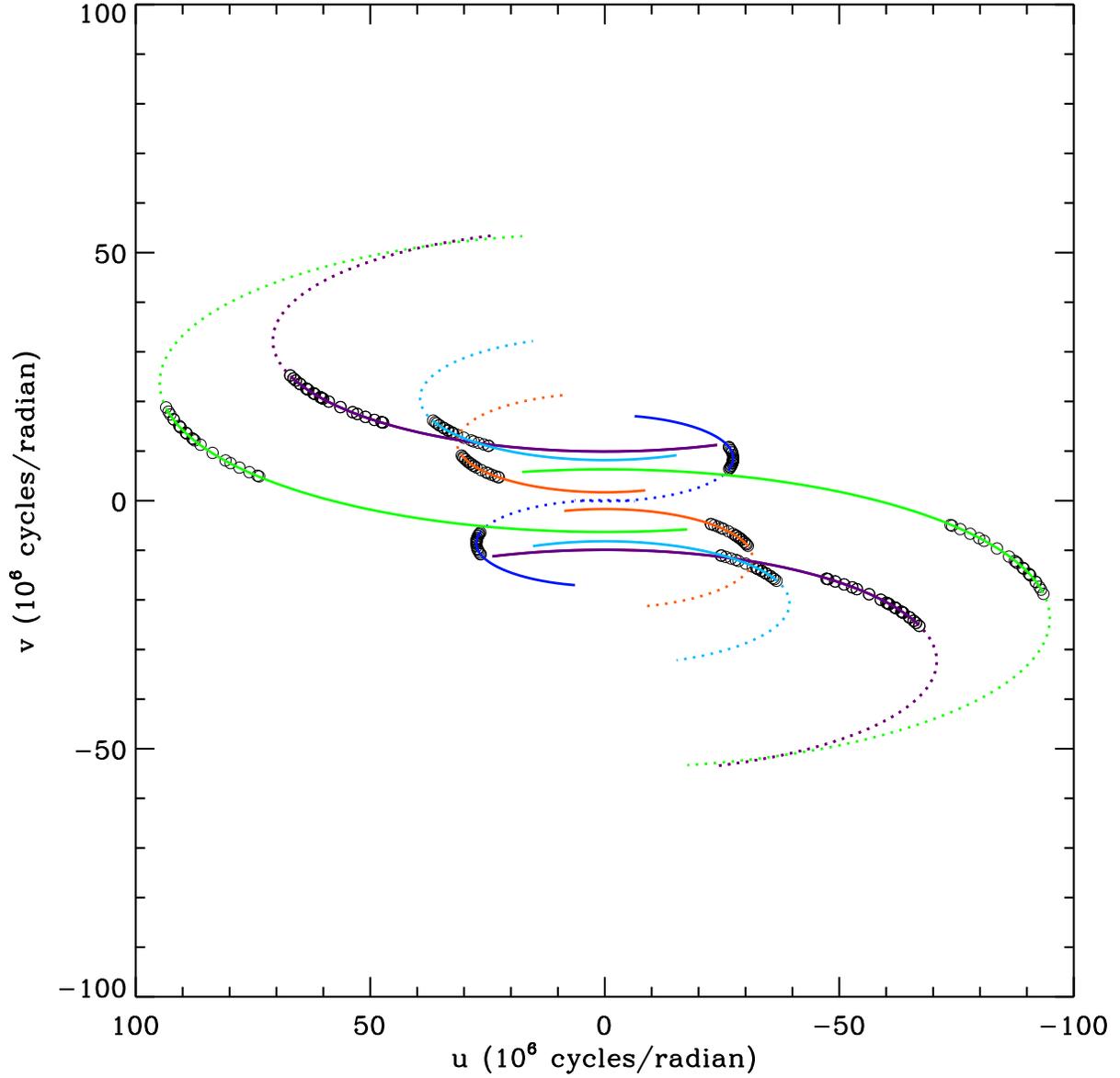}
\caption{The ($u$,$v$)-plane coverage of the interferometric
observations of $\chi$~Oph obtained in the H$\alpha$ channel with five
unique baselines from four nights of observations~({\it circles}).
The dotted-lines indicate possible coverage from meridian to 6 hr east
and the solid-lines indicate possible coverage from the meridian to 6
hr west.}
\label{fig:chi-oph-uv}
\end{figure}

We acquired interferometric observations of $\chi$~Oph on four nights
between 2006~June~11 and 2006~June~18.  All observations were made
with the same instrumental configuration. We used 4 different
telescopes, resulting in five unique baselines, with one baseline
signal measured twice on two independent output beams~\citep[see
Fig.~2 in][for a schematic of the beam combiner]{Tycner06b}. The
baselines used and their corresponding lengths are: AC--AE (18.9~m),
AC--AW (22.2~m), AW--W7 (29.5~m), AC--W7 (51.6~m), AE--W7 (64.4~m),
with the AC--W7 baseline measured on two output beams. The
observations were obtained using two spectrographs, each recording
simultaneously signal from 3 baselines at 16 spectral channels in the
wavelength range 560--870~nm. The resulting $(u,v)$-plane coverage of
the interferometric observations obtained from the spectral channel
containing the H$\alpha$ emission line is shown in
Figure~\ref{fig:chi-oph-uv}.

The observations of $\chi$~Oph were interleaved with observations of a
nearby calibrator star $\zeta$~Oph (O9V; HR~6175; HD~149757).  The
choice of the calibrator star was based on the relative proximity on
the sky to the target and the spectral type, which was similar to that
of the target star.  Furthermore, the angular diameter of $\zeta$~Oph
has been directly measured by~\citet{Hanbury74} using an intensity
interferometer with sufficient accuracy to be competitive with other
methods that estimate angular diameters based on theoretical or
empirical models.  Therefore, for the calibrator star, we adopted a
uniform disk angular diameter of 0.50$\pm$0.05~mas as obtained
by~\citet{Hanbury74}.

The data reductions followed standard NPOI procedures and will only
briefly be described here. The raw data are processed to produce
squared visibilities ($V^2$) averaged into 1~s intervals.  These data
are then flagged to eliminate points with fringe tracking and pointing
problems \citep{Hummel98,Hummel03}. The flagged data set is then used
to obtain 30~s averages~(we refer to these averages as scans), which
are bias corrected using data obtained off the fringe (these are also
known as incoherent scans).  Table~\ref{tab:obs} lists the individual
nights and the number of scans acquired on each night.  Lastly, the
squared visibilities from the spectral channel containing the
H$\alpha$ emission line are calibrated with respect to the continuum
channels~\citep[see][for a detailed description of the
procedures]{Tycner03, Tycner06a}.  Our final data set~(shown in
Fig.~\ref{fig:chi-oph-best-fit}) consists of the ensemble of all the
$V^2$ data from the H$\alpha$ channel obtained on all four nights
listed in Table~\ref{tab:obs}.  The individual calibrated H$\alpha$
$V^2$ values are also listed in Table~\ref{tab:v2data}.  Because
squared visibilities represent the normalized Fourier power of the
source structure on the sky, the interferometric observations can be
compared directly to the Fourier transform of the synthetic image
produced by the model.

The observations of $\chi$~Oph used in this study were not optimized
for the purpose of calibrating with respect to an external calibrator.
However, such a calibration allows one to inspect the continuum
channels for the signature of a resolved stellar disk or a binary
companion.  Using the scans obtained on $\zeta$~Oph to obtain the
instrumental system response, we found that large systematic residuals
from scan to scan remained after the calibration, which we attribute
to atmospheric variations on a timescale of minutes that are poorly
sampled by the scans on the calibrator.  For this reason, we can only
conclude that the central star at the continuum channels is marginally
resolved with a uniform disk diameter of $\lesssim$~1~mas.  An
interferometric signature of a resolved binary is less susceptible to
systematic variations from scan to scan because of its distinct
sinusoidal functional form that can be seen across different spectral
channels.  In fact, the NPOI has been used successfully to detect
binaries over a wide range of separations~\citep[see for
example][]{Hutter04}.  The search for a binary signature in
observations of $\chi$~Oph, however, did not yield any convincing
evidence of such a signature in our interferometric observations that
would correspond to a magnitude difference~($\Delta m$) of $\sim 3$ or
less at separations of few hundred milli-arcseconds~(mas) or less.
Although we cannot rule out a fainter companion, such a companion
would not contribute significantly to the total emission from the
system and therefore for the purpose of the analysis presented in this
study, $\chi$~Oph can be treated as a single star.  It should be
noted, however, that $\chi$~Oph has been classified as a single-lined
spectroscopic binary by more than one study~\citep{Abt78, Harmanec87,
Levato87}, although there is no agreement on the period of the binary,
which was estimated to be 138.8~d by~\citet{Abt78} and 34.1~d
by~\citet{Harmanec87}.

\subsection{Spectroscopy}

To estimate the strength of the H$\alpha$-emission in $\chi$~Oph at
the time of interferometric observations we obtained a high resolution
spectrum in the H$\alpha$ region using a fiber-fed Echelle
spectrograph at the Lowell Observatory's John S. Hall telescope.  The
spectrum was acquired on 2006~June~10, just one day before our one
week long interferometric run.  The spectroscopic data were processed
using standard routines developed specifically for the instrument used
to acquire the observations~\citep{Hall94}.  The final reduced
spectrum in the H$\alpha$ region reaches a resolving power of 10,000
and a signal-to-noise ratio of few hundred (see
Fig.~\ref{fig:Halpha}).

In addition to the spectrum shown in Figure~\ref{fig:Halpha} we have
also acquired spectroscopic observations of $\chi$~Oph in 2006 May and
2006 September.  The peak intensities and the equivalent widths (EWs)
of the H$\alpha$ emission line all agree to within 1\% with the line
profile shown in Figure~\ref{fig:Halpha}.  Therefore, we assume that
during our interferometric run the overall emission in the H$\alpha$
line was stable and combining all of the interferometric observations
into one data set is justified.

\section{Determining the Size of the H$\alpha$-emitting Region}
\label{sec:interferometry}

We followed the procedure described by \cite{Tycner06b} to determine
the angular extent of the H$\alpha$ disk. We fitted a circularly
symmetric Gaussian model to the interferometric data and obtained a
best-fit diameter (defined as full-width at half-maximum; FWHM) for
the disk of 3.46~$\pm$~0.07~mas, which at a distance of 150~pc
corresponds to a diameter of 0.52~AU~(112~$R_\sun$).  The fit results
in a reduced $\chi^2$ value of 0.95, and therefore, there is no
indication that the observational signature deviates from circular
symmetry or a Gaussian shape for that matter~(shown as dash-dotted
line in Fig.~\ref{fig:chi-oph-best-fit}).  In fact, this is the best
observational evidence to date that a Gaussian model is not only the
simplest mathematical description of the radial distribution of the
H$\alpha$ emission in a circumstellar disk, but it is also fully
consistent with the observed distribution.  This is in addition to
similar conclusions obtained previously by \citet{Tycner06b} for two
other Be stars, $\gamma$~Cas and $\phi$~Per.

It is interesting to compare our results based on the Gaussian model
fit with the simple estimator based on H$\alpha$ EW obtained by
\citet{Grundstrom06}.  With the stellar parameters for $\chi$~Oph
listed in Table~\ref{tab:parameters} and assuming that the disk is
viewed at 20$^{\circ}$~(see \S~\ref{sec:grid}), for an EW of $-7.1$~nm
based on our spectroscopic observations \citet{Grundstrom06}
predict\footnote{We have used the programs provided by the authors at
http://www.chara.gsu.edu/$\sim$gies/Idlpro/BeDisk.tar to extrapolate
for equivalent widths beyond the values considered in Fig.~1 of
\citet{Grundstrom06}.} a disk radius of 9.2~$R_{\star}$.  This agrees
well with the radius of $\approx 9.8 R_{\star}$ we obtain based on the
Gaussian fit~(using half of our FWHM measure).

Although characterizing the H$\alpha$ emitting region with a Gaussian
model can be a useful tool, especially when only a limited
interferometric data set is available to constrain the disk
characteristics, it is only a simple observational parameterization.
However, in cases where the interferometric data set covers wide range
of spatial frequencies it is desirable to extract more information
about the physical characteristics of the disk, such as density and
temperature distribution.  For this purpose a much more sophisticated
model is needed.

\section{Disk Model}
\label{sec:disk}

The radiative equilibrium model used to represent the thermal
structure of the gaseous circumstellar disk surrounding $\chi\,$Oph
has been computed using the {\sc bedisk} code, which is described in
detail by~\citet{Sigut07}.  This code incorporates many improvements
over previous treatments, most notably the use of a solar chemical
composition for the circumstellar gas, which is an important
ingredient for calculating heating and cooling rates. These models are
ideally suited to compare with interferometric observations because
model monochromatic, 2-dimensional images of the circumstellar disk at
specific wavelengths can be computed. The {\sc bedisk} code can also
be used to compute the hydrogen line spectra and the overall spectral
energy distribution. For the calculation of the hydrogen line
profiles, the disk is assumed to be in pure Keplerian rotation. There
is considerable evidence that Be star disks are indeed rotationally
supported based on the detailed analysis of line profiles
\citep{hum00b} and on the interpretation of V/R variations as
one-armed density waves in the disk, which are predicted for Keplerian
disks \citep{oka07}.  Recent work based on long-baseline
interferometry has also provided observational evidence in support of
Keplerian rotation~\citep{Meilland07}.

The disk density model of~\citet{Sigut07} requires only a few input
stellar and disk parameters. It is assumed that the density in the
equatorial plane of the disk is given by an $R^{-n}$ power-law in
radial distance (where $R$ is the distance from the star's rotation
axis), and that the density distribution vertical to the equatorial
plane (in the $Z$ direction) is set by the hydrostatic equilibrium
established by the gas pressure gradient and the vertical component of
the star's gravitational acceleration.  Such a model generally
produces a thin disk in which $Z/R\ll 1$, and in this case the form of
the density distribution is particularly simple:
\begin{equation}
\label{eq:rhoTo}
\rho(R,Z) = \rho_0 \left(\frac{R_{\star}}{R}\right)^{n}
e^{-\left(\frac{Z}{H}\right)^2} ,
\end{equation}
where $\rho_0$ is the density at the inner edge of the disk in the
equatorial plane, $n$ is the index in the radial power-law, and $H$ is
the scale height in the $Z$ direction and is given by
\begin{equation}
H = \sqrt{\frac{2R^3}{\alpha_0}},
\end{equation}
with the parameter $\alpha_0$ of the form:
\begin{equation}
\alpha_0 = GM_{\star} \frac{\mu_0 m_{\rm H}}{kT_0} .
\end{equation}
In these expressions, $M_{\star}$ and $R_{\star}$ are the stellar mass
and radius, respectively; $T_0$ and $\mu_0$ are the assumed
(vertically isothermal) temperature and mean-molecular weight at the
radial distance $R$. This simple, analytical form for the density is
possible because it is assumed that the vertical pressure scale height
at each $R$ can be represented by a single temperature $T_0$.
Although this assumption might not be directly applicable to dense
disks where strong vertical temperature gradients in the disk might be
present, \citet{SMJ07} examined the accuracy of this assumption in
detail (in pure hydrogen disks) and found that in most cases, the
differences in typical predicted diagnostics, such as the H$\alpha$
profiles and IR excesses, are not large for models in consistent
radiative and (vertical) hydrostatic equilibrium if the
density-averaged disk temperature is used for the parameter $T_0$. The
largest variations are found for the densest models (with largest
$\rho_0$) as these models develop a cool, equatorial zone close to the
star. Hence for this current work, we have adopted the simpler density
structure given by equation~\ref{eq:rhoTo} with the parameter $T_0$
being constant for all values of $R$ and $Z$.  In the case of models
constructed for $\chi$~Oph we set $T_0$ at 10,500~K based on typical
density-weighted disk temperatures obtained for models in reasonable
agreement with the observations of $\chi$~Oph.

With this model for the disk gas density the parameters $\rho_0$ and
$n$ are free to be varied to match the interferometric visibilities
and H$\alpha$ line profile of $\chi\;$Oph.  Before we can obtain a
self-consistent solution for a disk model with specific $\rho_0$ and
$n$ values, we need to adopt the parameters for the central star.  The
reported $T_{\rm eff}$ values for $\chi$~Oph in the literature range
from the low of 18,000~K to a high of
29,600~K~\citep{Goraya84,Waters86,Fremat05, Zorec05}.  Furthermore,
\citet{Goraya84} not only detected variations in visual magnitude of
$\chi$~Oph but also reported that such variations were accompanied by
changes in the slope of the continuum, resulting in variations of
derived $T_{\rm eff}$ values by more than 4,000~K.  Because we do not
attempt to model the intrinsic variability of the source in this
study, we adopt $T_{\rm eff}$ based on the spectral type of
$\chi$~Oph, a main sequence B2 star, which based on tabulations
of~\citet{deJager87} gives a $T_{\rm eff}$ of 20,900~K.  The stellar
mass and radius are less critical parameters for the disk models and
we adopted the average values based on the spectral type tabulated by
\citet{AQ} of $10.9 M_{\sun}$ and $5.7 R_{\sun}$.  The model input
parameters for the stellar component are listed in
Table~\ref{tab:parameters}.

Figure~\ref{fig:tempplot} shows the thermal structure of a disk model
with $n=2.5$ and $\rho_0 = 2 \times 10^{-11}$ g cm$^{-3}$ as a
function of radial distance from the central star and distance from
the equatorial plane.  Although the computational grid extends out to
285~$R_{\sun}$~(50 $R_{\star}$) from the central star, the figure
shows only the region out to 200~$R_{\sun}$.  There is a conspicuous
cooler region in the equatorial plane of the disk within the first
$\sim 50 R_{\sun}$ from the central star.  This cooler volume
represents the region of the disk where the density and the
corresponding optical depths are larger.  The disk temperature reaches
a minimum of $\approx 6,000$~K in this volume. The density
perpendicular to the equatorial plane falls off approximately
exponentially, so regions surrounding this cool volume have
significantly lower densities and are optically thin resulting in
greater temperatures.  These features in the disk thermal structure
are typical of other models presented in the literature for Be stars
with moderate to high disk densities \citep[for example see][]{jon04,
Carciofi06, Sigut07}.  The overall average temperature of the disk
over the computation grid is $\approx 10,400$~K, which is consistent
with our adopted temperature used in the vertical hydrostatic
equilibrium calculations (recall $T_0$).  It should also be noted that
the temperature structure shown in Figure~\ref{fig:tempplot} is
influenced by the incoming radiation from the central star and
therefore ultimately the thermal structure will be affected by the
choice of stellar parameters, such as $T_{\rm eff}$ and $R_{\star}$.

\section{Constraining the Model}

\subsection{Synthetic H$\alpha$ Image}
\label{sec:image}

A self-consistent temperature solution for the disk density model
(like that shown in Fig.~\ref{fig:tempplot}) can be used to directly
compute the specific intensity of radiation emitted perpendicular to
the plane of the disk. As the hydrogen level populations are found
naturally as part of the thermal solution, the H$\alpha$ line
emissivity and opacity,
\begin{equation}
\label{eq:emission}
\eta^{32}_{\nu} = \frac{h\nu_{23}}{4\pi} N_3\,A_{32}\,\phi_{\nu} \,,
\end{equation}
and
\begin{equation}
\label{eq:absorption}
\chi^{23}_{\nu} = \frac{h\nu_{23}}{4\pi} (N_2\,B_{23}-N_3\,B_{32})\,\phi_{\nu} \,,
\end{equation}
are known at each $(R,Z)$ location in the disk. Here $N_2$ and $N_3$
are the hydrogen number densities in levels 2 and 3 respectively,
$B_{23}$, $B_{32}$ and $A_{32}$ are the Einstein probability
coefficients for H$\alpha$, and $\phi_{\nu}$ is the H$\alpha$ line
profile. For the latter, we have adopted the routines of
\citet{Barklem03} which include the contributions of (thermal) Doppler
broadening and collisional broadening (including the linear Stark
effect). Turbulent (or microturbulent) broadening was not considered.

To form the total opacity and emissivity at the frequency of
H$\alpha$, we have added the continuous free-free and bound-free
emissivity and opacity of hydrogen to equations~\ref{eq:emission}
and~\ref{eq:absorption}.  We have also included electron scattering as
a coherent source of opacity but not as a source of emissivity. In a
radiative equilibrium (or thermal-balance) calculation it is not usual
to include continuum scattering as it leads to no net energy gain or
loss by the gas.  However, non-coherent electron scattering is capable
of broadening spectral lines and blurring continuum
edges~\citep{Rybicki94}, and this process could potentially effect the
H$\alpha$ EW measure.  Although these effects are ignored in the
current study, non-coherent electron scattering will be included in
future work.  Given the opacity and emissivity, we solve the equation
of radiative transfer at each radial distance $R$ in the disk along a
ray in the $Z$-direction, i.e.,
\begin{equation}
\frac{dI_{\nu}(R,Z)}{dZ} = \eta^{32}_{\nu} - \chi^{23}_{\nu}\,I_{\nu}(R,Z) \,.
\end{equation}
The formal solution was performed with the short-characteristics
method of \citet{Olson87}. This gives the specific intensity emerging
perpendicular to the disk and is what an external observer would see
if the disk was viewed at $i=0^{\circ}$.

To obtain a profile that can be compared with observations, the
kinematic broadening of the disk's (assumed) Keplerian rotation must
be included.  As we estimate that the inclination of the rotation axis
of $\chi\,$Oph is quite small ($i\approx\,20^{\circ}$, see below), we
have assumed that the specific intensity at this small inclination is
well approximated by the $i=0^{\circ}$ intensities computed
above. Given this, we simply shift the $i=0^{\circ}$ spectrum of each
projected area of the disk and star on the sky by its radial velocity.
For areas on the stellar surface, we have adopted the photospheric
H$\alpha$ profiles computed in LTE by the code of
\citet{Barklem03}. The radial velocity of each patch on the stellar
surface followed from the assumption of a spherical star in solid body
rotation with an equatorial velocity of 375~km~s$^{-1}$ (giving a
$v\sin i$ of 144~km~s$^{-1}$; see below).  Clearly this procedure is
not appropriate for large inclination angles as the perpendicular rays
through the disk would no longer be representative of the rays in the
direction towards the external observer. However, direct comparison of
the above procedure with a complete routine that solves the transfer
equation along the inclined rays pointing at the observer shows that
the simpler procedure gives good results for small inclinations and
the approximate procedure is computationally much faster.

We specify more than a hundred wavelength steps to cover the 15~nm
wide spectral region centered at the H$\alpha$ line, which corresponds
to the spectral channel used to obtain the interferometric
observations.  The integrated net intensity over that 15~nm spectral
window is shown in Figure~\ref{fig:radial-intensity} where clearly the
total intensity is dominated by the bright central star.  This is
expected because the 15~nm wide spectral region is much wider than the
H$\alpha$ emission line~(recall Fig.~\ref{fig:Halpha}) and therefore
the H$\alpha$ channel receives significant fraction of the light from
the central star.  In turn, this radial intensity distribution can be
used to construct a circularly symmetric image that corresponds to the
total intensity in the H$\alpha$ channel.

\begin{figure}
\plotone{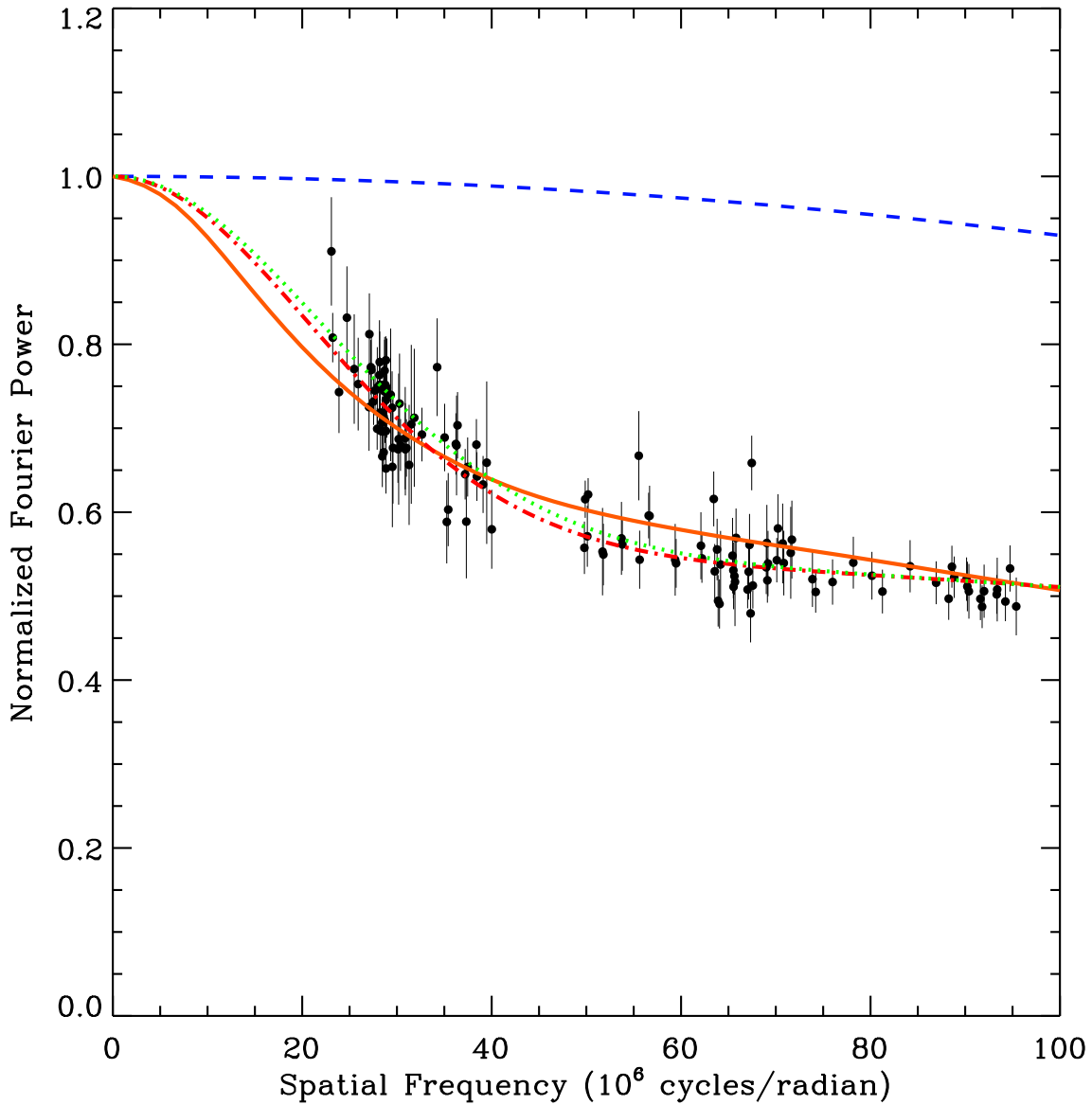}
\caption{ Interferometric data obtained from the spectral channel
containing the H$\alpha$ emission line from the circumstellar disk of
$\chi$~Oph.  The 0.52~AU diameter (FWHM) circumstellar disk is fully
resolved, whereas the central star is assumed to have an uniform disk
diameter of 0.35~mas ({\it dashed line}).  The best-fit model~({\it
solid line}) obtained by taking the Fourier transform of the synthetic
image from Figure~\ref{fig:image} is shown along with a circularly
symmetric Gaussian model with FWHM diameter of 3.46~mas ({\it
dash-dotted line}).  The effect of an axial ratio of 0.94 on the
Gaussian model is also shown ({\it dotted line}). }
\label{fig:chi-oph-best-fit}
\end{figure}

\begin{figure}
\plotone{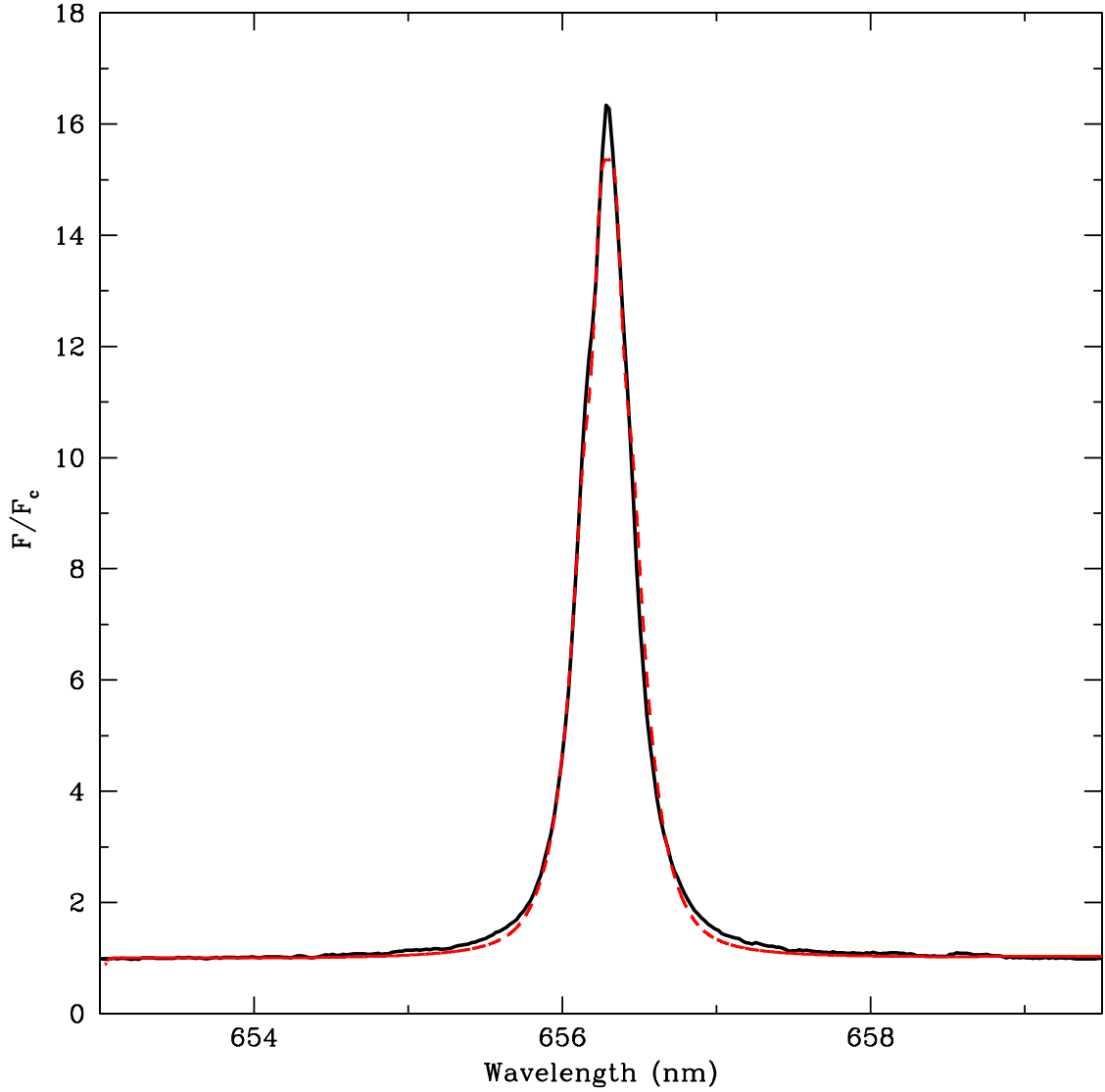}
\caption{The H$\alpha$ emission line of $\chi$~Oph observed on
2006~June~10 ({\it solid-line}) and the synthetic H$\alpha$ profile
calculated from the best-fit disk model ({\it dashed-line}).  The
model profile has been obtained for an inclination angle $i$ of
20$^{\circ}$ and has been broadened by a Gaussian kernel to match the
resolving power~($R$ = 10,000) of the spectroscopic observations.  The
observed H$\alpha$ profile has an EW of $-7.1\pm 0.2$~nm and the
theoretical profile has EW of $-6.6$~nm.}
\label{fig:Halpha}
\end{figure}

\begin{figure}
\plotone{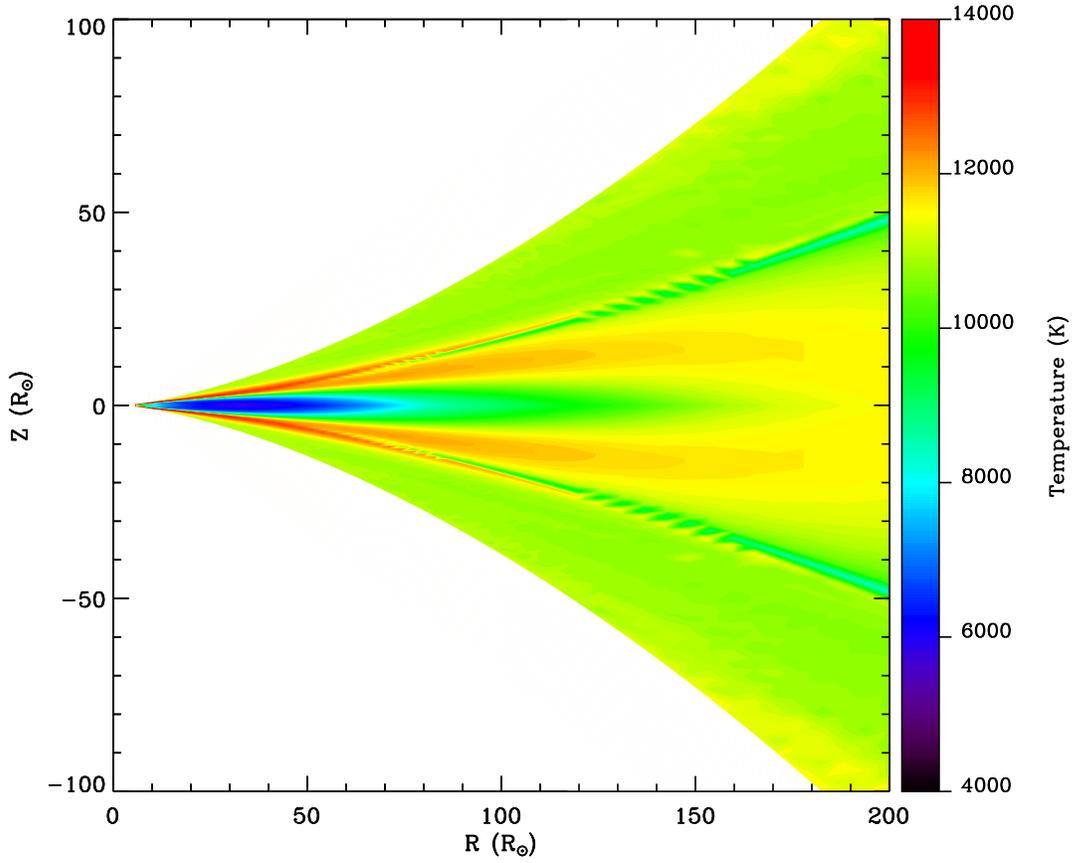}
\caption{The thermal structure of the disk model of $\chi$~Oph
obtained using $n = 2.5$, $\rho_0 = 2.0 \times 10^{-11}$ g cm$^{-3}$
and the stellar parameters listed in Table~\ref{tab:parameters}.  The
thermal structure is shown as a function of radial distance~($R$) from
the center of the star and the vertical distance~($Z$) above and below
the equatorial plane. }
\label{fig:tempplot}
\end{figure}

\begin{figure}
\plotone{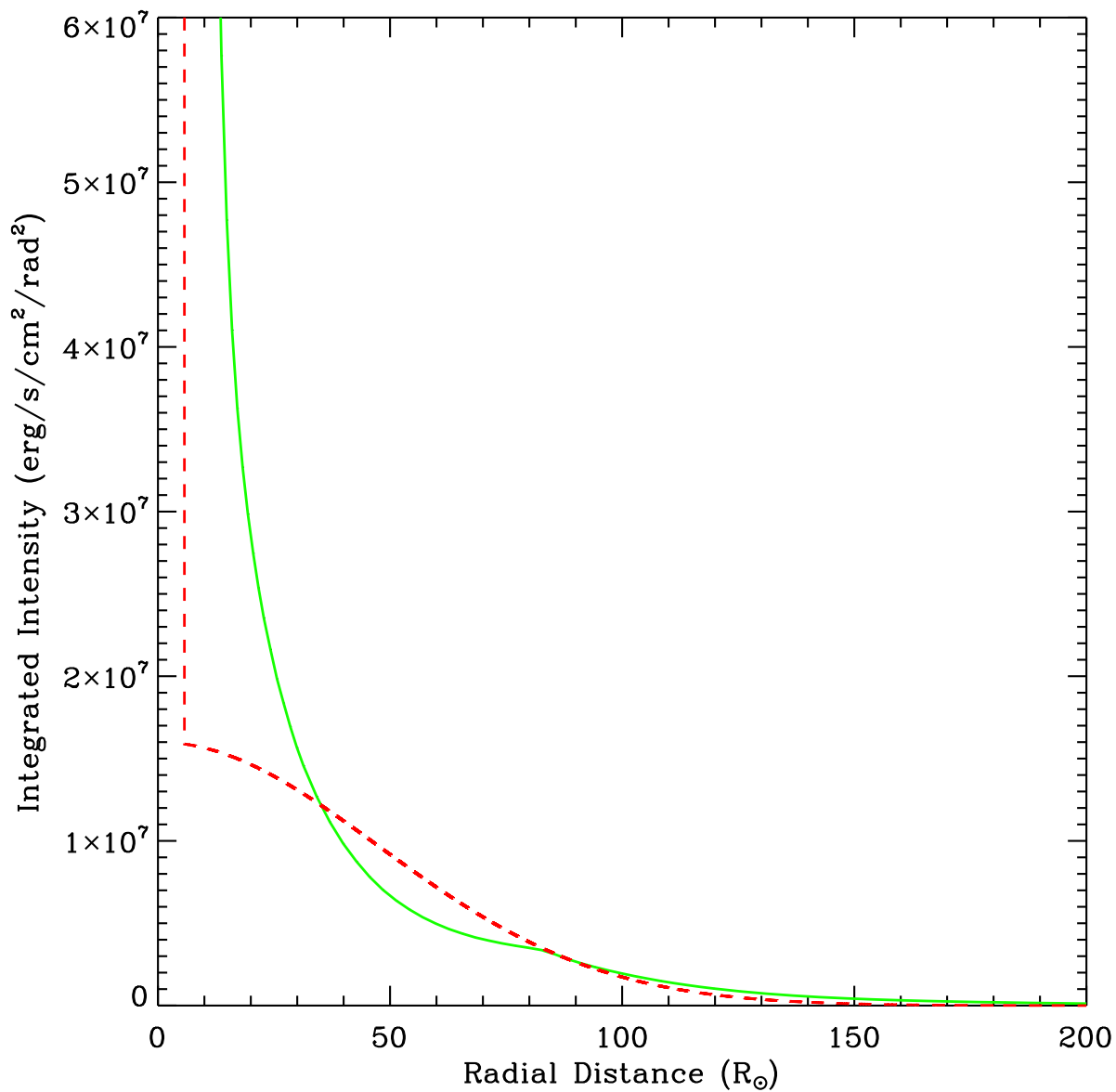}
\caption{Integrated model intensity for the direction normal to the
plane of the disk in a 15~nm wide spectral region centered at
H$\alpha$ ({\it solid line}) for the best-fit model of $n = 2.5$ and
$\rho_0 = 2.0 \times 10^{-11}$ g cm$^{-3}$.  The Gaussian radial
intensity distribution that reproduces the interferometric signature
in Fig.~\ref{fig:chi-oph-best-fit} is also shown~({\it dashed line}).}
\label{fig:radial-intensity}
\end{figure}

The relatively small $v \sin i$ value reported for $\chi$~Oph of
144~km~s$^{-1}$ by \citet{Zorec05} and the lack of interferometric
signature that would suggest large deviations from circular
symmetry~(recall \S\ref{sec:interferometry}), both suggest that the
inclination angle of the disk to the plane of the sky is small.
Assuming that the star is rotating near its critical velocity of
477~km~s$^{-1}$~\citep[estimated based on its spectral type; see][and
references therein]{Porter96}, the inclination angle $i$ must be $\sim
18^{\circ}$ or more for the star not to rotate above its critical
velocity.  As rapid rotation is a known characteristic of Be stars, we
adopt 20$^{\circ}$ for the inclination angle of $\chi$~Oph.
Furthermore, because the axial ratio produced by the projection effect
on geometrically thin and circularly symmetric disks scales as a
cosine of the inclination angle, an inclination of 20$^{\circ}$
produces only a $\sim$6\% departure from circular symmetry (i.e., an
axial ratio of 0.94).  For circumstellar disks with non-negligible
opening angles the effect would be even smaller.  However, an axial
ratio of 0.94 would be undetectable in our data because it affects the
$V^2$ values at a level much smaller than the precision of our
interferometric observations~(compare the dash-dotted and dotted
curves in Fig.~\ref{fig:chi-oph-best-fit}).  Therefore, for the
purpose of this study we will treat the synthetic images as circularly
symmetric and will ignore the small projection effect on the images.

\begin{figure}
\plotone{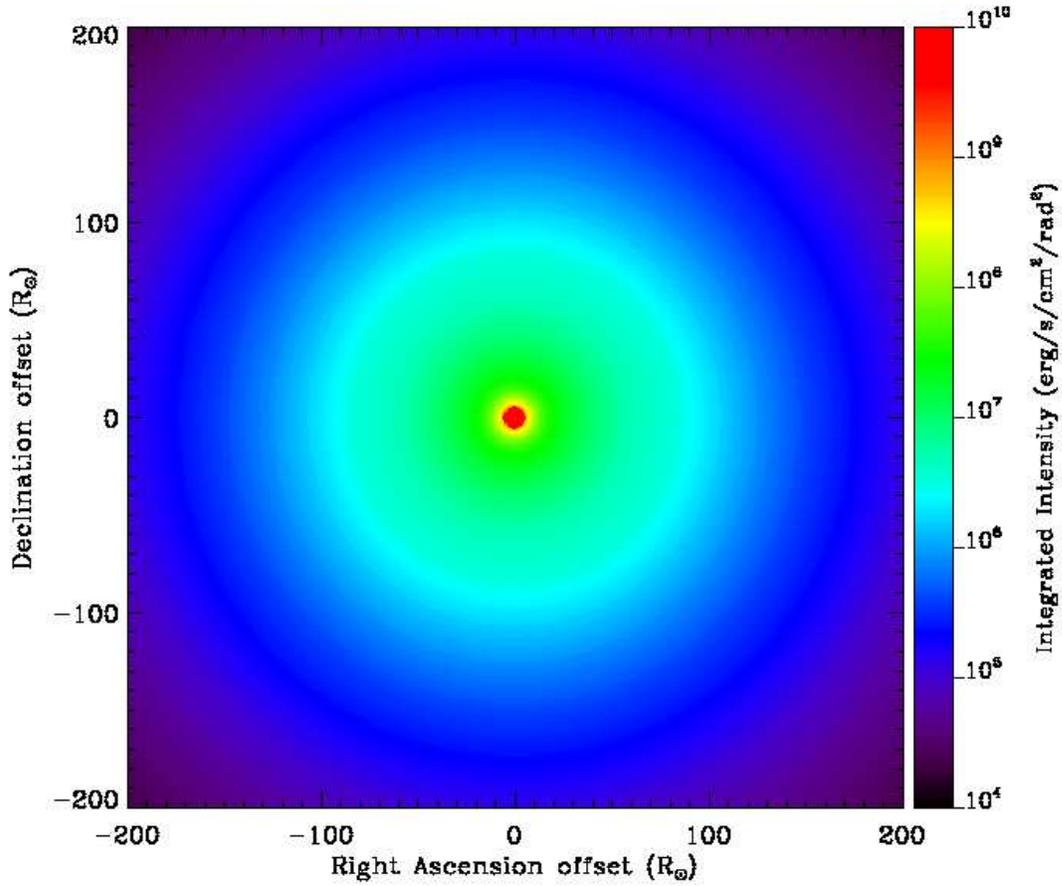}
\caption{Synthetic image of the best-fit disk model viewed pole on.
Only intensities in a 15~nm spectral window centered at H$\alpha$
emission line were used to construct this synthetic image.  The
surface brightness of the central star is derived using the stellar
atmosphere model integrated over the same 15~nm spectral window as
used to calculate the net intensity from the disk.}
\label{fig:image}
\end{figure}

The squared visibilities from the H$\alpha$ channel measure the
normalized Fourier power of the source structure on the sky, and
therefore the Fourier transform of the model image and the
interferometric observations can be directly compared.  To accomplish
this we first construct a circularly symmetric image corresponding to
the radial distribution of the integrated intensity over the 15~nm
wide spectral bandpass covered by the NPOI spectral channel containing
the H$\alpha$ emission line.  Because the model intensities are
calculated for radial region extending up to 50~$R_{\star}$ from the
central star, we require minimum dimensions of the model image of
100~$R_{\star}$ $\times$ 100~$R_{\star}$ to fully describe the output.
However, we extend our synthetic image to 1000~$R_{\star}$ $\times$
1000~$R_{\star}$ dimensions (the outer regions are filled with zeros)
to increase the sampling frequency of the model in the Fourier space.
The image is sampled every 0.2~$R_{\star}$ and this is sufficient to
avoid any aliasing problems at the high spatial frequencies covered by
the observations.  The region of interest of the synthetic image is
shown in Figure~\ref{fig:image}.  The 2-D Fourier transform of this
circularly symmetric image results in a function that also has
circular symmetry.  The normalized Fourier power of this transform can
then be plotted as a function of the radial spatial frequency~(shown
as a solid line in Fig.~\ref{fig:chi-oph-best-fit}).  We use the
standard $\chi^2$ statistic to assess the goodness-of-fit of the model
to the actual data obtained on the source.

\subsection{The Parameter Space}
\label{sec:grid}

\begin{figure}
\plotone{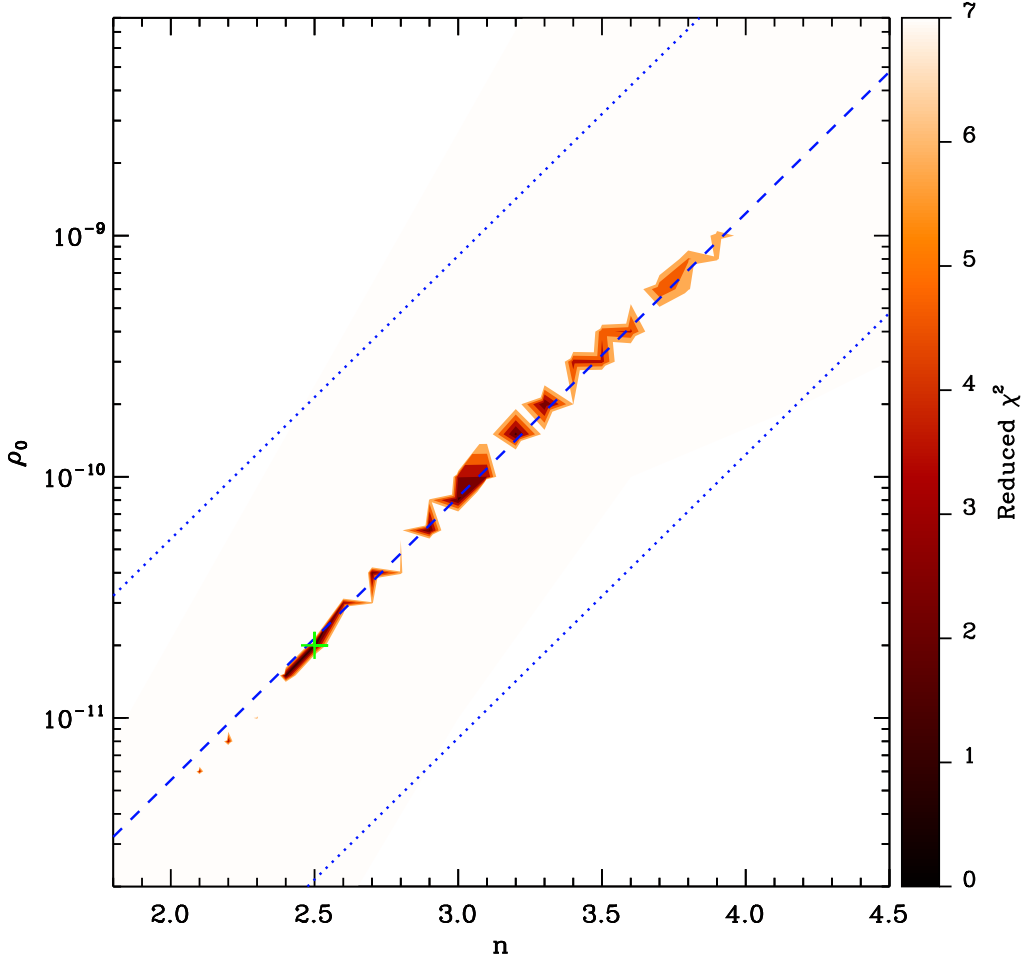}
\caption{Contour plot showing the reduced $\chi^2$ value as a function
of $\rho_0$~(g cm$^{-3}$) and $n$.  Only a region of reduced $\chi^2 <
7$ is shown and the best-fit model with reduced $\chi^2$ of 1.17 is
marked with a plus sign.  Some of the gaps in the linear trend are
caused by the grid spacing and the limitations of the contour plotting
routine dealing with unequal grid spacing.  The approximate boundary
of the region of parameter space sampled by our models is also
indicated~({\it dotted lines}), along with a linear fit~({\it dashed
line}) to the models with reduced $\chi^2 < 7$ in the log~$\rho_0$ and
$n$ plane~(see \S~\ref{sec:grid}).  }
\label{fig:chi2}
\end{figure}

We explore the parametric region of of the disk density
model~(eq.~\ref{eq:rhoTo}) by computing over 500 parameter pairs of
$n$ and $\rho_0$.  We covered the range of 1.8 -- 5.3 in $n$ and
adjusted $\rho_0$ from a low of $1.0 \times 10^{-12}$~g~cm$^{-3}$ to a
high of $8.0 \times 10^{-8}$~g~cm$^{-3}$.  The majority of the
solutions were completely inconsistent with the interferometric
observations, and those solutions that yielded reduced $\chi^2$ value
of 7 or less were concentrated in a very well defined 'valley' in the
$\chi^2$ space~(see Fig.~\ref{fig:chi2}).

The model that results in the lowest reduced $\chi^2$ value of 1.17
corresponds to $n$ of 2.5 and $\rho_0$ of $2.0 \times
10^{-11}$~g~cm$^{-3}$.  The thermal
structure~(Fig.~\ref{fig:tempplot}), H$\alpha$ model
image~(Fig.~\ref{fig:image}), and the model curve shown in
Figure~\ref{fig:chi-oph-best-fit} all correspond to this best-fit
model.  Furthermore, assuming that the disk is viewed at an
inclination angle of 20$^{\circ}$~(as discussed in
\S~\ref{sec:image}), we obtain an excellent agreement with the
observed H$\alpha$ line profile obtained through spectroscopy~(see
Fig.~\ref{fig:Halpha}).  In fact, for the disk model defined by our
best-fit to interferometric data, the H$\alpha$ profile can only be
reproduced assuming $i$ of 20$^{\circ}$.  For any smaller inclination
the model profile is simply too sharp~(due to the decreased rotational
broadening) and for larger values of $i$ the line becomes too broad
and weaker than the actual observed profile.

In addition to the best-fit solution found at $n$ of 2.5 and $\rho_0$
of $2.0 \times 10^{-11}$~g~cm$^{-3}$ it is clearly evident from
Figure~\ref{fig:chi2} that there exists a range of parameter values in
the $\chi^2$ space that produce disk models that fit the observational
data acceptably well.  However, these solutions appear to be confined
to a very narrow range in the $\chi^2$ space.  We attribute this range
of solutions to the changing total H$\alpha$ flux generated by a
model, where most of the solutions in the upper-left part of the
figure produce too much H$\alpha$ flux with respect to the central
star and those in the lower-right part of the figure do not produce
enough H$\alpha$ flux.  These effects can be related to different disk
thermal structures and different densities, which in turn will affect
the apparent sizes of the H$\alpha$-emitting regions.  A linear
least-squares fit to solutions with reduced $\chi^2 < 7$ in the
log~$\rho_0$ versus $n$ plane yields a slope of $1.18\pm 0.03$ (and an
intercept of $-13.6\pm 0.1$), where the uncertainty does not account
for the effects of gridding or the choice of the reduced $\chi^2$
cutoff value.  Nevertheless, this indicates that at least for a range
of $n$ values between 2.5 and 4.0 the disk solutions that produce
interferometric signatures closest to the observational data fall
along a relation where log~$\rho_0$ is directly proportional to $n$.
This is in agreement with similar conclusions made by \citet{Gies07}
who approximate the size of the disk on the sky using a boundary
between the optically thick and thin disk regions, which then leads to
the expectation that good fits fall along a relation where
log~$\rho_0$ is proportional to $n$.

\section{Discussion}

It is useful to compare our best-fit disk model for $\chi$~Oph to
previous results of \citet{Waters86} who modeled the IRAS IR excesses
of a large sample of Be stars with disk models with fixed opening
angles of $15^{\circ}$ and where the density varied only with the
distance from the stellar center, $r$, in the form of
\begin{equation}
\label{eq:rhorens}
\rho(r) = \rho_0' (r/R_{\star})^{-n} .
\end{equation}
Assuming an isothermal disk with a temperature of 18,000~K, a
temperature that is considerably larger than the $\sim$10,400~K
density-weighted average temperature of our best-fit model,
\citet{Waters86} found a radial power-law index $n$ of 2.4 and a
density at the base of the disk of $\log \rho_0=-11.4$.  These results
compare well to our results of $n=2.5$ and $\log\rho_0=-10.7$, even
though the disk-density models used in the two studies are not the
same.  The smaller base disk density found by \citet{Waters86} is
likely the result of the fact that his density distribution is
constant along radial arcs over the entire opening angle of the disk
which places more material in the disk as compared to our models.

\citet{Porter99} modeled the circumstellar regions of Be stars as
isothermal viscous disks and predicted $n$ of $7/2$.  Viscous models
that drop the assumption of an isothermal disk can find a range of $n$
values as the disk fills or empties \citep{JSP08}.  \citet{Porter99}
also analyzed the IR excess of $\chi$~Oph using the data of
\citet{Waters86}, but with a disk model density essentially equivalent
to ours. In the case of an isothermal disk (also set to 18,000~K),
\citet{Porter99} found an index $n$ of 2.2 with $\log \rho_0 =-11.2$;
in the case of a disk in which the density-averaged temperature was
allowed to vary with radial distance as a power-law, \citet{Porter99}
obtained $n=1.9$ and $\log(\rho_0)=-11.2$.  Porter indicates that the
non-isothermal disk formally fits the IR data best, but that the
improvement is not large.

\begin{figure}
\plotone{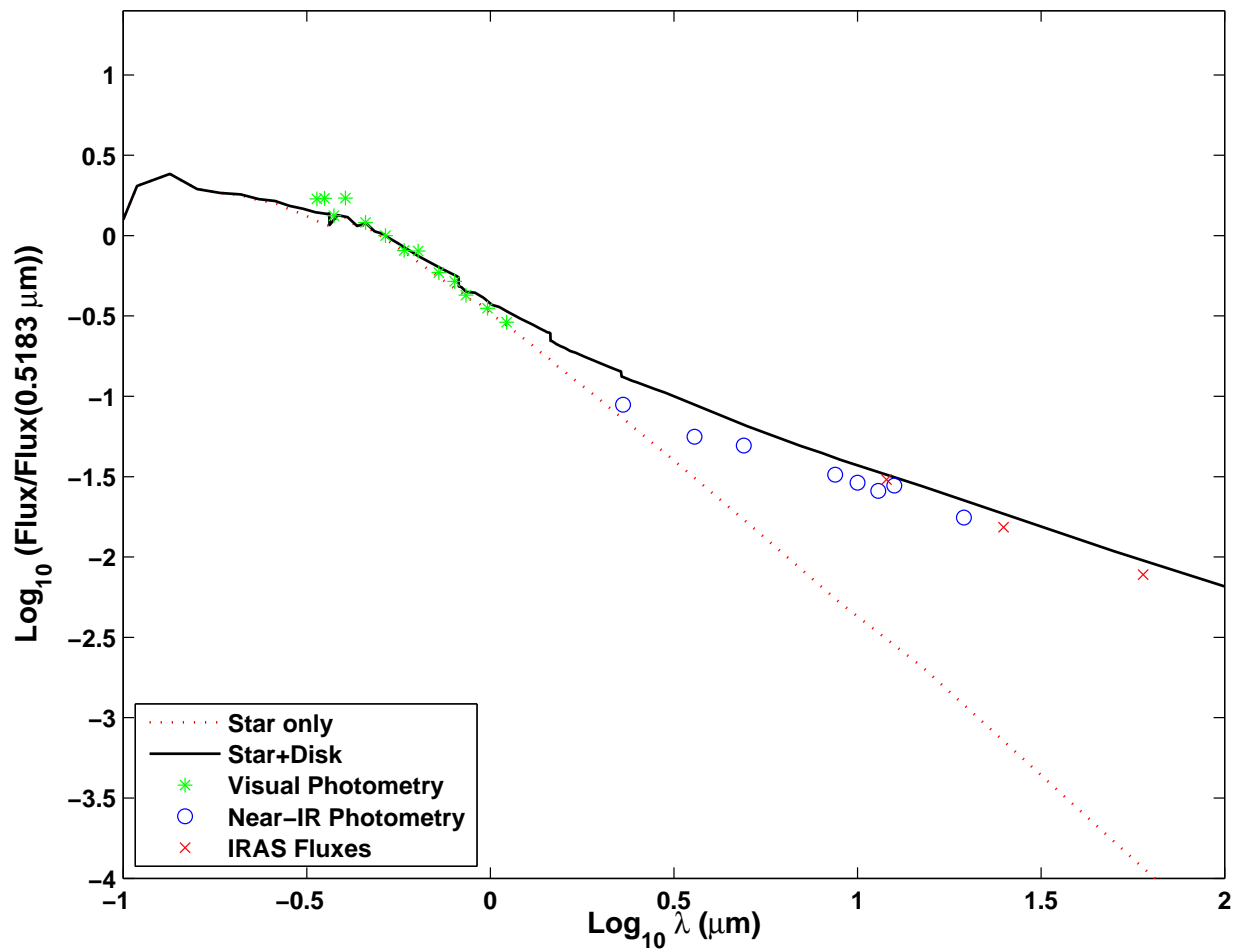}
\caption{ The spectral energy distribution (SED) of our best-fit model
corresponding to $n = 2.5$ and $\rho_0 = 2.0 \times 10^{-11}$ g
cm$^{-3}$ (solid-line), which consists of the sum of the stellar
contribution~(dotted-line) and IR excess produced by the circumstellar
disk.  The model SED is compared to visual, near-IR and far-IR
observations reported in literature \citep[see][and references
therein]{Waters86}. }
\label{fig:IR-fit}
\end{figure}

In addition to the comparison between model and observed H$\alpha$
interferometric visibilities and the H$\alpha$ line profile, it is
also instructive to compare the predicted IR excess of our best-fit
model to the IR observations reported in the literature.
Figure~\ref{fig:IR-fit} illustrates the comparison between the SED of
our best-fit model and the visual and IR observations from
\citet{Waters86}.  The overall agreement is quite reasonable,
especially considering that the model was not fit to the SED data,
although the observations in the $1-10\,\mu$m region fall below the
model.  We should also point out that variability between the
observations used by \citet{Waters86} and our 2006 H$\alpha$
observations (interferometric and spectroscopic) cannot be ruled out.
For example, \citet{ban00} obtained the H$\alpha$ profile for
$\chi$~Oph in 1998 and reported an equivalent width of approximately
$-3.6$~nm, as compared to the $-7.1\pm0.2$~nm found in the present
work for the 2006 epoch~(recall Fig.~\ref{fig:Halpha}). Hence, the
H$\alpha$ emission of $\chi\,$Oph has approximately doubled in 8
years, and assuming that the H$\alpha$ emission was similarly weaker
at the time of observations of \citet{Waters86} we can then expect our
best-fit model to over-predict IR excess as compared to Waters'
results.  For this reason, we feel that our model cannot be improved
using the IR observations that were acquired using the IRAS satellite
in the early 1980's.

\section{Summary and Future Work}

In this paper we presented the study of the circumstellar disk of the
Be star $\chi$~Oph, based on the approach of combining high-spatial
resolution interferometric observations with numerical disk models
requiring only a few input parameters. Using the interferometric data
we were also able to determine that the H$\alpha$-emitting disk can be
fit by a circular symmetric Gaussian with FWHM diameter of
3.46$\pm$0.07~mas.  Using the code {\sc bedisk}~\citep{Sigut07} we
created a grid of models with a range of inner edge densities
($\rho_0$) and equatorial plane density distribution fall-off rates
($n$), while keeping the mass, radius and effective temperature of the
central star fixed.  These models were used to create synthetic
H$\alpha$ images by adding the emission line and continuum flux,
including the contribution from the central star, in a 15~nm spectral
region around H$\alpha$, similar to that of the H$\alpha$
interferometric channel. These images were then Fourier transformed
and compared directly to the interferometric observations, resulting
in the best-fit model with $n = 2.5$ and $\rho_0 = 2.0 \times
10^{-11}$~g~cm$^{-3}$.  The best-fit disk model was also used to
calculate a synthetic H$\alpha$ line profile and spectral energy
distribution of this system over 0.1--100~$\mu$m, which show good
agreement with the observed line profile and the photometric
observations published in the literature.

The technique demonstrated here, of combining models with
interferometric observations, demonstrates a new and independent
method for obtaining properties of circumstellar disks of Be stars.
This is possible because the density distribution in the disk is
strongly affected by the value of the disk density at the stellar
surface, $\rho_0$, and to a lesser extent by $n$.  Variations in
$\rho_0$ and $n$ result in dramatic changes in the thermal structure
of the disk.  Ultimately, the density distribution and the resulting
thermal structure directly affect the H$\alpha$ flux predicted by the
theoretical models, which in turn can be constrained by observations
that spatially resolve the region.  In the future, we plan to
investigate more sophisticated forms of the model density distribution
within Be star disks, such as rotating density perturbation models
\citep[see for example][]{wis07}, and density distributions that are
consistent with hydrodynamic simulations.

\acknowledgements 

The Navy Prototype Optical Interferometer is a joint project of the
Naval Research Laboratory and the U.S. Naval Observatory, in
cooperation with Lowell Observatory, and is funded by the Office of
Naval Research and the Oceanographer of the Navy.  We thank Doug Gies
for the very helpful suggestions on how to improve this manuscript.
C.~T. thanks Lowell Observatory for the generous telescope time
allocation on the John S. Hall Telescope.  C.~T. would also like to
thank Erika Grundstrom for useful discussions and thanks Nick Melena
who contributed to the reductions of interferometric observations.
C.~E.~J. and T.~A.~A.~S. would like to acknowledge support from the
Natural Sciences and Engineering Research Council of Canada.

{\it Facilities:} \facility{NPOI}, \facility{LO:42in}




\clearpage

\begin{table}[htp]
\caption[]{\sc \small NPOI Observations of $\chi$~Oph}
\label{tab:obs}
\begin{tabular}{lcc} \hline\hline
\hspace{2cm} UT Date \hspace{2cm}      & \# of Scans  &  \# of Baselines \\ \hline
2006 June 11               \dotfill &       5      &         5        \\
2006 June 13               \dotfill &       4      &         5        \\
2006 June 17               \dotfill &       6      &         5        \\
2006 June 18               \dotfill &       7      &         5        \\
\hline
\end{tabular}
\end{table}

\begin{table}[htp]
\caption[]{\sc \small Calibrated H$\alpha$ Squared Visibilities of $\chi$~Oph}
\label{tab:v2data}
\begin{tabular}{crrcc} \hline\hline
 Julian Date        & Spatial Frequency $u$ & Spatial Frequency $v$ &               &         \\
(JD $-$ 2,450,000) & ($10^6$ cycles/radian)  & ($10^6$ cycles/radian)  &   $V_{\rm{H}\alpha}^2$ & Baseline$^{\dag}$ \\ \hline
3897.785  &  $  27.003$  &  $  -7.122$  &  0.749 $\pm$ 0.048 & AC-AE \\
3897.785  &  $  64.880$  &  $  23.478$  &  0.534 $\pm$ 0.032 & W7-AC \\
3897.785  &  $  91.883$  &  $  16.357$  &  0.502 $\pm$ 0.023 & W7-AE \\
3897.785  &  $ -29.723$  &  $  -8.345$  &  0.688 $\pm$ 0.061 & AC-AW \\
3897.785  &  $  64.989$  &  $  23.518$  &  0.519 $\pm$ 0.027 & W7-AC \\
3897.785  &  $  35.266$  &  $  15.173$  &  0.680 $\pm$ 0.030 & W7-AW \\
3897.801  &  $  27.324$  &  $  -7.970$  &  0.696 $\pm$ 0.042 & AC-AE \\
3897.801  &  $  61.798$  &  $  21.501$  &  0.548 $\pm$ 0.045 & W7-AC \\
3897.801  &  $  89.123$  &  $  13.531$  &  0.520 $\pm$ 0.026 & W7-AE \\
3897.801  &  $ -28.537$  &  $  -7.436$  &  0.724 $\pm$ 0.043 & AC-AW \\
3897.801  &  $  61.902$  &  $  21.537$  &  0.511 $\pm$ 0.027 & W7-AC \\
3897.801  &  $  33.365$  &  $  14.101$  &  0.681 $\pm$ 0.037 & W7-AW \\
\hline
\end{tabular}\\[0.9ex]
\parbox{6.0in}{\footnotesize \quad {\sc Note.} ---
Table~\ref{tab:v2data} is published in its entirety in the electronic
edition of the Astrophysical Journal.  $^{\dag}$ There are 6 baselines
per scan (same JD) with the W7--AC measured at two output beams with
slightly different wavelength scales and thus resulting in slightly
different $u$ and $v$ values for the same baseline length. }
\end{table}

\begin{table}[htp]
\caption[]{\sc \small Model Stellar Parameters}
\label{tab:parameters}
\begin{tabular}{lccl} \hline\hline
\hspace{1.7cm} Parameter \hspace{1.7cm}      & Symbol  &     Value     &    Reference \\ \hline
Mass ($M_\odot$)                \dotfill &  $M_{\star}$        &  10.9                    &  \citet{AQ}  \\
Radius ($R_\odot$)              \dotfill &  $R_{\star}$        &  5.7                    &  \citet{AQ} \\
Effective temperature (K)       \dotfill &  $T_{\rm eff}$      &  20,900                 &  \citet{deJager87} \\
Luminosity ($L_\odot$)          \dotfill &  $L_{\star}$        &  $5.6\times 10^3$     &  calculated$^{\dag}$ \\
Surface gravity                  \dotfill &  $\log g$          &  $4.0$      &  calculated$^{\ddag}$ \\
\hline
\end{tabular}\\[0.9ex]
\parbox{6.0in}{\footnotesize \quad {\sc Note.} --- $^{\dag}$ $L_\star
= 4\pi R_{\star}^2 \sigma T_{\rm eff}^4$ where $\sigma$ is the
Stefan-Boltzmann constant. $^{\ddag}$ $\log g = \log (GM_\star
R_\star^{-2})$ where $G$ is the gravitational constant. }
\end{table}

\clearpage


\begin{thebibliography}{}
\bibitem[Abt \& Levy(1978)]{Abt78} Abt, H.~A., \& Levy, S.~G.\ 1978,
\apjs, 36, 241
\bibitem[Banerjee et al.(2000)]{ban00} Banerjee, D.\ P.\ K., Rawat, S.\ D., \&
Janardhan, P.\ 2000, A\&AS 147, 229
\bibitem[Barklem \& Piskunov(2003)]{Barklem03} Barklem, P.~S., \&
Piskunov, N.\ 2003, Modeling of Stellar Atmospheres, 210, 28P
\bibitem[Carciofi \& Bjorkman(2006)]{Carciofi06} Carciofi, A.~C., \&
Bjorkman, J.~E.\ 2006, \apj, 639, 1081
\bibitem[Cot\'{e} \& Waters(1987)]{Cote87} Cot\'{e}, J., \& Waters,
L. B. F. M. 1987, \aap, 176, 93
\bibitem[Cox(2000)]{AQ} Cox, A.~N.\ 2000, Allen's astrophysical
quantities, 4th ed.~Publisher: New York: AIP Press; Springer, 2000
\bibitem[de Jager \& Nieuwenhuijzen(1987)]{deJager87} de Jager, C., \&
Nieuwenhuijzen, H.\ 1987, \aap, 177, 217
\bibitem[Dougherty(1994)]{dou94} Dougherty, S.M.\ et al., 1994, A\&A
290, 609
\bibitem[Fr{\'e}mat et al.(2005)]{Fremat05} Fr{\'e}mat, Y., Zorec, J.,
Hubert, A.-M., \& Floquet, M.\ 2005, \aap, 440, 305
\bibitem[Gehrz et al.(1974)]{Gehrz74} Gehrz, R.~D., Hackwell, J.~A.,
\& Jones, T.~W.\ 1974, \apj, 191, 675
\bibitem[Gies et al.(2007)]{Gies07} Gies, D.~R., et al.\ 2007, \apj,
654, 527
\bibitem[Goraya(1984)]{Goraya84} Goraya, P.~S.\ 1984, \aap, 138, 19
\bibitem[Grundstrom \& Gies(2006)]{Grundstrom06} Grundstrom, E.~D., \&
Gies, D.~R.\ 2006, \apjl, 651, L53
\bibitem[Hall et al.(1994)]{Hall94} Hall, J. C., Fulton, E. E.,
Huenemoerder, D. P., Welty, A. D., \& Neff, J. E. 1994, \pasp, 106,
315
\bibitem[Hanbury Brown et al.(1974)]{Hanbury74} Hanbury Brown, R.,
Davis, J., \& Allen, L.~R.\ 1974, \mnras, 167, 121
\bibitem[Harmanec(1987)]{Harmanec87} Harmanec, P.\ 1987, Bulletin of
the Astronomical Institutes of Czechoslovakia, 38, 283
\bibitem[Hummel et al.(1998)]{Hummel98} Hummel, C. A., Mozurkewich, D., Armstrong,
J. T., Hajian, a. R., Elias II, N. M., \& Hutter, D. J. 1998, \aj, 116, 2536
\bibitem[Hummel(2000)]{hum00a} Hummel, W., in {\it The Be Phenomena in
Early-Type Stars}, M.\ A.\ Smith, H.\ F.\ Henrichs, \& J.\ Fabregat
(eds), 2000, ASP Conf.\ Ser.\ 214, 396
\bibitem[Hummel \& Vrancken(2000)]{hum00b} Hummel, W., \& Vrancken, M
2000, \aap, 359, 1075
\bibitem[Hummel et al.(2003)]{Hummel03} Hummel, C. A., et al. 2003a,
\aj, 125, 2630
\bibitem[Hutter et al.(2004)]{Hutter04} Hutter, D.~J., Benson, J.~A.,
Zavala, R.~T., Johnston, K.~J., Pauls, T.~A., Hummel, C.~A., \&
Armstrong, J.~T.\ 2004, \procspie, 5491, 73
\bibitem[Jones, Sigut \& Marlborough(2004)]{jon04} Jones, C. E.,
Sigut, T. A. A., \& Marlborough, J. M. 2004, \mnras, 352, 841
\bibitem[Jones et al.(2008)]{JSP08} Jones, C.\ E., Sigut, T.\ A.\ A.,
\& Porter, J.\ M.\ 2008, \mnras, 386, 1922
\bibitem[Kastner \& Mazzali(1989)]{Kastner89} Kastner, J.~H., \&
Mazzali, P.~A.\ 1989, \aap, 210, 295
\bibitem[Levato et al.(1987)]{Levato87} Levato, H., Malaroda, S.,
Morrell, N., \& Solivella, G.\ 1987, \apjs, 64, 487
\bibitem[Marlborough(1969)]{Marlborough69} Marlborough, J.~M.\ 1969,
\apj, 156, 135
\bibitem[Meilland et al.(2007)]{Meilland07} Meilland, A., et al.\
2007, \aap, 464, 59
\bibitem[Millar \& Marlborough(1998)]{Millar98} Millar, C. E., \&
Marlborough, J. M. 1998, \apj, 494, 715, MM
\bibitem[Millar \& Marlborough(1999)]{Millar99} Millar, C.~E., \&
Marlborough, J.~M.\ 1999, \apj, 526, 400
\bibitem[Okazaki(2007)]{oka07} Okazaki, A.~T.\ 2007, Active OB-Stars:
Laboratories for Stellare and Circumstellar Physics, Edited by
S. Stefl, S. P. Owocki, and A. T. Okazaki., 361, 230
\bibitem[Olson \& Kunasz(1987)]{Olson87} Olson, G.~L., \& Kunasz,
P.~B.\ 1987, Journal of Quantitative Spectroscopy and Radiative
Transfer, 38, 325
\bibitem[Perryman et al.(1997)]{Perryman97} Perryman, M.~A.~C., et
al.\ 1997, \aap, 323, L49
\bibitem[Porter(1996)]{Porter96} Porter, J.~M.\ 1996, \mnras, 280, L31
\bibitem[Porter(1999)]{Porter99} Porter, J.~M.\ 1999, \aap, 348, 512
\bibitem[Quirrenbach et al.(1997)]{Quirrenbach97} Quirrenbach, A., et 
al.\ 1997, \apj, 479, 477 
\bibitem[Rybicki \& Hummer(1994)]{Rybicki94} Rybicki, G.~B.,
\& Hummer, D.~G.\ 1994, \aap, 290, 553
\bibitem[Sigut \& Jones(2007)]{Sigut07} Sigut, T. A. A. \& Jones,
C. E.  2007, \apj, 668, 481
\bibitem[Sigut et al.(2007)]{SMJ07} Sigut, T.\ A.\ A., McGill, M.\ A.,
  \& Jones, C.\ E.\ 2007, ApJ, submitted
\bibitem[Tycner et al.(2003)]{Tycner03} Tycner, C., Hajian, A. R.,
Mozurkewich, D., Armstrong, J. T., Benson, J. A., Gilbreath, G. C.,
Hutter, D. J., Pauls, T. A., \& Lester, J. B.  2003, \aj, 125, 3378
\bibitem[Tycner et al.(2005)]{Tycner05}Tycner, C. et al. 2005, \apj,
624, 359
\bibitem[Tycner et al.(2006a)]{Tycner06a} Tycner, C., Benson, J.~A.,
Hutter, D.~J., Schmitt, H.~R., \& Zavala, R.~T.\ 2006a, \procspie,
6268, 49
\bibitem[Tycner et al.(2006b)]{Tycner06b} Tycner, C., et al.\ 2006b,
\aj, 131, 2710
\bibitem[Waters(1986)]{Waters86} Waters, L. B. F. M. 1986 \aap, 162, 121
\bibitem[Waters et al.(1987)]{wat87} Waters, L. B. F. M., Cote, J., \&
Lamers, H. J. G. L. M. 1987, \aap, 185, 206
\bibitem[Wisniewski et al.(2007)]{wis07} Wisniewski, J. P., Kowalski,
A. F., Bjorkman, K. S., Bjorkman, J. E., \& Carciofi, A. C. 2007,
\apj, 656, L21
\bibitem[Wood et al.(1997)]{woo97} Wood K., Bjorkman K. S., \&
Bjorkman J. E. 1997, \apj, 477, 926
\bibitem[Zorec et al.(2005)]{Zorec05} Zorec, J., Fr{\'e}mat, Y., \&
Cidale, L.\ 2005, \aap, 441, 235
\end{thebibliography}
\end{document}